\documentclass[amssymb,amsmath,pra,twocolumn,floatfix,showpacs]{revtex4}
\usepackage{graphicx}

\begin{document}
\title{Maximizing the Hilbert space for a finite number of distinguishable quantum states.}
\author{Andrew D. Greentree$^1$, S. G. Schirmer$^2$, F. Green$^1$,
Lloyd C. L. Hollenberg$^3$, A. R. Hamilton$^{1}$, and R. G.
Clark$^{1}$}

\affiliation{$^1$ Centre for Quantum Computer Technology, School
of Physics, The University of New South Wales, Sydney, NSW 2052,
Australia.}

\affiliation{$^2$ Dept of Applied Maths + Theoretical Physics
(DAMTP) and Dept of Engineering, University of Cambridge,
Cambridge, CB2 1PZ, UK.}

\affiliation{$^3$ Centre for Quantum Computer Technology School of
Physics, University of Melbourne, Vic. 3010, Australia.}

\date{\today}

\begin{abstract}
We consider a quantum system with a finite number of
distinguishable quantum states, which may be partitioned freely by
a number of quantum particles, assumed to be maximally entangled.
We show that if we partition the system into a number of qudits,
then the Hilbert space dimension is maximized when each quantum
particle is allowed to represent a qudit of order $e$.  We
demonstrate that the dimensionality of an entangled system,
constrained by the total number quantum states, partitioned so as
to maximize the number of qutrits will always exceed the
dimensionality of other qudit partitioning.  We then show that if
we relax the requirement of partitioning the system into qudits,
but instead let the particles exist in any given state, that the
Hilbert space dimension is greatly increased.
\end{abstract}


\maketitle

Quantum computation \cite{bib:NielsenBook} has, in remarkably
short time, become one of the most interesting fields in applied
quantum physics today. There are numerous suggestions for
implementing quantum computers.  One common feature of all
scalable quantum computers, is their use of entangled particles
\cite{bib:Blume-Kohout2002}.

Most quantum computer proposals focus on qubits (quantum bits) as
the fundamental element of computing.  These are two state quantum
systems that can be entangled. Further work has begun to examine
the possibility of performing computations with qudits (quantum
digits) \cite{bib:QuditRef}.  Qudits are an extension of qubits,
that are systems with any (integer) number of states greater than
1. The qubit is therefore the two-state qudit, and the qutrit is
the three-state qudit.  For convenience we will refer to the
$x$-state qudit as an x-qudit.

Investigations into qudits have shown many significant results.
Entanglement between two qutrits was first discussed by Caves and
Milburn \cite{bib:QutritEntanglement}.  Bell inequalities for
systems of qudits are more strongly violated than analogous
systems of qubits \cite{bib:BellViolation}, and recent experiments
have shown entanglement of two qutrits, realized using the orbital
angular momentum as the quantum state \cite{bib:VaziriPRL2002}. A
quantum-communication protocol using qutrits has also been
proposed \cite{bib:QutritCommunication}. Universal quantum gates
and gate fidelity have been studied for qudit systems
\cite{bib:QuditGates}, a readout method based on quantum state
tomography has been proposed \cite{bib:QST} and the extension of
infinite-dimensional qudits to continuous-variable quantum
computation has also been made \cite{bib:SandersPreprint2002}.

The Hilbert-space dimension of a quantum computer has been
identified as the primary resource for any physical realization of
such a device \cite{bib:Blume-Kohout2002}.  To that end it is
necessary to consider how best to maximize the Hilbert-space
dimension for a given quantum computer geometry. It is obvious
that for a given number of particles, the dimensionality of the
Hilbert space, ${\cal D}$, will be increased by increasing the
dimensionality of the particles, although it is not, in general,
trivial to increase the dimensionality of the fundamental
particles being used for the computation; nor is the added
complexity involved necessarily favorable.  However there are
proposals for quantum computing which involve a number of quantum
particles having available a finite number of distinguishable
quantum states. Examples include the charge-qubit scheme
\cite{bib:EkertRMP1996,bib:Hollenberg}, superconducting
Cooper-pair boxes \cite{bib:CooperPairBox}, quantum computing
based on photons in interferometers joined by lossless nonlinear
optical elements \cite{bib:QNLO} and the linear optics
implementation \cite{bib:KLM}. In all these proposals the
fundamental quantum particles (electrons, Cooper-pairs or photons)
may be in a finite number of distinguishable quantum states
(positions, interferometer arms).  These quantum states are
grouped together and we term such groupings \textit{quantum
elements}.  For qudits we have one particle per quantum element,
which can be in a superposition of all states within the element.
Ideally every particle will be entangled with every other
particle.

\begin{figure}[tb!]
\includegraphics[width=0.9\columnwidth,clip]{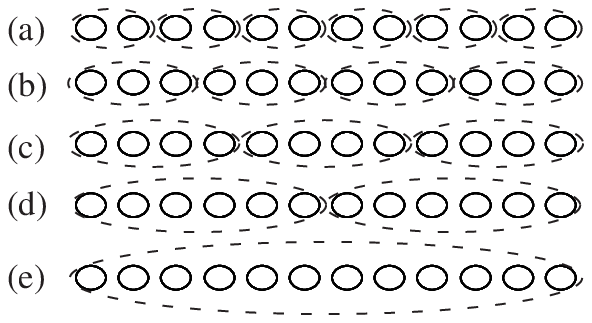}
\caption{\label{fig:Schematic} Possible groupings of a linear
array of 12 distinguishable quantum sites into various quantum
qudit partitions.  Qudit groupings are shown schematically as
dashed ovals, and open circles are quantum sites. (a) Partitioning
into six qubits, (b) partitioning into four qutrits, (c)
partitioning into three 4-qudits, (d) partitioning into two
6-qudits, (e) partitioning into one 12-qudit.}
\end{figure}

For concreteness, we will consider the charge scheme of Hollenberg
\textit{et al.} \cite{bib:Hollenberg}, although the generalization
to other schemes is straightforward. Without examining issues
related to operational complexity or physical architecture we will
show ways in which to optimize the Hilbert-space dimensionality
for a system where the \textit{number of accessible quantum
states} is unchanging, although the logical groupings of those
states into quantum elements, and the number of particles shared
between those states is varied.

Consider a linear array of $N$ donor impurities, with controllable
tunnelling probabilities and individual energy levels.  If we
partition the space into $x$-qudits, as shown in Fig.
\ref{fig:Schematic} for $x=2,3,4$, then we must introduce $N/x$
electrons.  The dimension of the Hilbert space obtained by
maximally entangling these N/x electrons is therefore:
\begin{eqnarray}
{\cal D}_x = x^{N/x}. \label{eq:Dimensionality}
\end{eqnarray}
Simply differentiating Eq. \ref{eq:Dimensionality} allows us to
determine the value of $x$ which maximizes ${\cal D}$ for a given
$N$
\begin{eqnarray*}
\frac{d{\cal D}}{dx} = -N x^{\frac{N}{x} - 2}(\ln x - 1) = 0, \\
\Rightarrow x = e.
\end{eqnarray*}
As each particle needs to have an integer number of levels, we
note that for identical qudits, the dimensionality is optimized
for qutrits.  The Hilbert-space dimensionality for qutrits over
qubits will be larger by a factor ${\cal D}_3/{\cal D}_2 =
\exp\left[ N\left( \ln 3 /3 - \ln 2 /2 \right)\right]$.

\begin{figure}[tb!]
\includegraphics[width=0.9\columnwidth,clip]{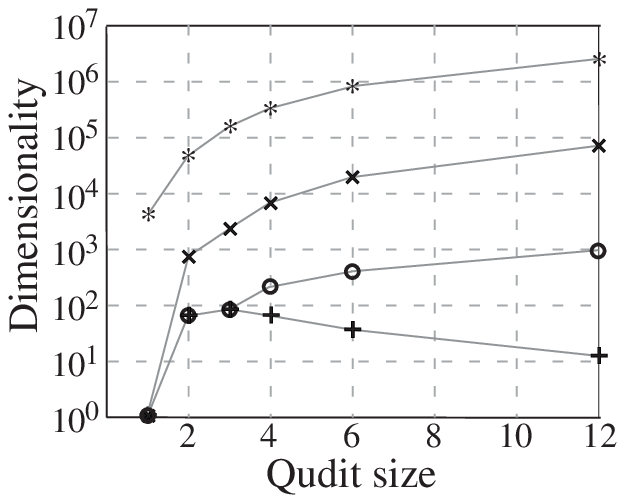}
\caption{\label{fig:CompareDim} Hilbert space dimensionality for
various geometries as a function of qudit size for $N$
distinguishable quantum states. $+$ corresponds to
equipartitioning into qudits, $\circ$ to a $Q_x(x/2,1)$ partition,
$\times$ to a $Q_x(x,2)$ partition and $\ast$ to spin referenced
systems. Note that the dimensionalities are only valid when $N/x$
is an integer greater than one, i.e. for $x = 2,3,4,6,12$.  The
lines are merely guides for the eye.}
\end{figure}

In Fig. \ref{fig:CompareDim} we show the Hilbert-space
dimensionality of a twelve-state system as a function of the size
of each quantum element.  Each different case is explained through
the text.  The $+$ symbols in Fig. \ref{fig:CompareDim} show the
dimensionality associated with qudit arrangements.  As expected,
the dimensionality is maximized by a qudit size of 3 (note that we
are only showing integer qudit sizes).  For the system with 12
quantum states, ${\cal D}_2 = 64$ whereas ${\cal D}_3 = 81$.

Conceivably, it may be possible to realize a quantum computer
which uses a mixture of qubits and qutrits so as to optimize the
total dimensionality of the Hilbert space, as recently discussed
by Daboul {\it et al.} \cite{bib:DaboulPreprint2002}. For
instance, suppose we partition the space into $y_2$ qubits and
$y_3$ qutrits such that $2y_2+3y_3=N'$ where $N' \lesssim N$.  One
might guess that the dimensionality of the Hilbert space would be
increased by mixing qubits and qutrits such that the average
dimensionality of the system $(2y_2+3y_3)/(y_2+y_3)$ is closer to
the optimal value $e$. However, this is not the case.  To see why,
note that the dimensionality of the Hilbert space of a system
partitioned as suggested is given by $2^{y_2} 3^{y_3} =
2^{(N'-3y_3)/2} 3^{y_3}$. With ${\cal D}_e = exp(N'/e)$ this leads
to
\begin{eqnarray*}
{\cal D}/{\cal D}_e = \exp(m y_3 + b),
\end{eqnarray*}
where $m=\ln(3)-\frac{3}{2}\ln(2)>0$ and
$b=(\frac{\ln(2)}{2}-e^{-1})N'$.  Note that $\exp(m y_3+b)=1$ if
$y_3=-b/m \approx 0.36 N'$.  Since $y_3\le N'/3$ on physical
grounds, this implies $my_3+b<0$, i.e., the exponential is always
less than one.  Furthermore, since $m>0$, the value of the
exponential \emph{increases} with the number of qutrits $y_3$.
Thus, the dimensionality of the Hilbert space is optimized if we
partition the system into the largest available number of qutrits,
except where adding a single qubit would more efficiently use all
of the available quantum states.

\begin{figure}[tb!]
\includegraphics[width=0.9\columnwidth,clip]{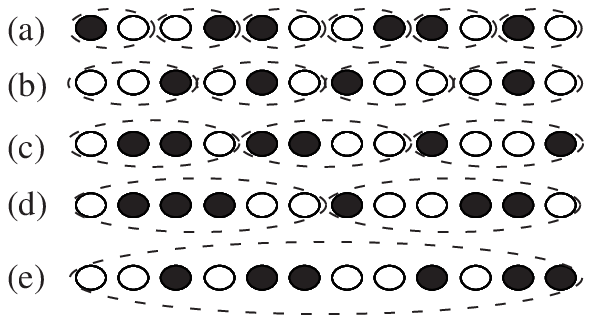}
\caption{\label{fig:PackingSchematic} Possible groupings of a
linear array of 12 distinguishable quantum sites into various
$Q_x(k,1)$ partitions.  The x-qudit groupings are shown
schematically as dashed ovals, open circles are quantum sites and
filled circles correspond to (randomly chosen) filled sites. (a)
six $Q_2(1,1)$ partitions, (b) four $Q_3(1,1)$ partitions, (c)
three $Q_4(2,1)$ partitions, (d) two $Q_6(3,1)$ partitions, (e)
one $Q_{12}(6,1)$ partition.}
\end{figure}

Allowing the number of particles to vary within each quantum
element changes the system dimensionality and generalizes the
qudit scheme.  We consider quantum elements which are groupings of
$x$ quantum states (modes), with $k$ particles per quantum element
and at most $l$ particles per state.  Such an arrangement is shown
schematically in Fig. \ref{fig:PackingSchematic}.  We term these
structures $(k,l)$ packing x-qudits and use the shorthand notation
$Q_x(k,l)$, writing the dimensionality of these structures as
$\dim[Q_x(k,l)]$. In essence, this approach reverses the work of
Abrams and Lloyd \cite{bib:AbramsPRL1997} where an algorithm for
efficient simulation of many-body Fermi systems on a quantum
computer was presented, by proposing a many-body system as a
quantum computer in its own right, and is in keeping with Bravyi
and Kitaev's work on fermionic quantum computation
\cite{bib:BravyiPreprint2000}. Blume-Kohout  \textit{et al.}
\cite{bib:Blume-Kohout2002} calculated the dimensionality of
Fermionic and Bosonic systems, which in our notation would
correspond to $Q_x(k,1)$ and $Q_x(k,k)$ respectively.  We note
that analysis of two electrons in a polygonal dot ($Q_x(2,1)$),
has been performed by Creffield \textit{et al.}
\cite{bib:Creffield}. Creffield and Platero studied two electrons
in a square dot \cite{bib:CreffieldPRB2002}, considering the
possibility of double occupancy ($Q_4(2,2)$).  This system
(without double occupancy) has also been proposed as a scalable
quantum element by Jefferson \textit{et al.}
\cite{bib:JeffersonPRA2002} ($Q_4(2,1)$).

We first consider the case that there can be no more than one
particle per quantum state, but the number of particles per
element is chosen to maximise the Hilbert-space dimensionality,
i.e. the $Q_x(k,1)$ partition, or Fermionic system.   In order to
maximize the total Hilbert-space dimensionality, it is clear that
the number of particles per quantum element must be equal to the
number of sites divided by 2, i.e. $k=x/2$ for $x$ even, or $k =
(x \pm 1)/2$ for $x$ odd, where the $\pm$ is due to symmetry
between particles and holes.  The total dimensionality for a
system with $N$ sites, partitioned into $N/x$ elements will
therefore be (assuming $x$ even and $N/x$ an integer)
\begin{eqnarray*}
{\cal D} = \{\dim [Q_x(x/2,1)]\}^{N/x} = {x \choose {x/2}}^{N/x}.
\end{eqnarray*}
where ${n \choose m}$ is the number of possible identifiably
different permutations of $n$ elements of two different types, $m$
of which are of type 1 (eg. particles) and $n-m$ of which are of
type 2 (eg. holes or empty sites). The Hilbert-space
dimensionality showing this form of partitioning is represented by
the open circles $\circ$ in Fig. \ref{fig:CompareDim}.  The total
dimensionality, $\dim[Q_x(x/2,1)]$, is clearly maximized when $x =
N$, which corresponds to the case that all particles are allowed
to roam freely in the lattice, which is the case of local
fermionic modes discussed by Bravyi and Kitaev
\cite{bib:BravyiPreprint2000} and is scalable, as discussed by
Blume-Kohout \cite{bib:Blume-Kohout2002}.  In a twelve state
configuration, for example, Fig. \ref{fig:PackingSchematic}(e)
shows the configuration which maximizes the dimensionality of the
Hilbert-space, with $\dim [Q_{12}(6,1)] = 924$, an order of
magnitude larger than case of qutrits. Despite this maximization,
it is important to realize that there may well be advantages to
partitioning into smaller subspaces (i.e. $x<N$) - for example it
is not possible to do error correction without some degree of
redundancy in coding.

The next obvious extension is to allow the maximum number of
particles per site to increase to two.  This could physically
correspond to an electron based quantum computer, where the
charging energy of two particles per dot can be overcome
\cite{bib:CreffieldPRB2002}. In this case the dimensionality of
the system will be
\begin{eqnarray*}
{\cal D} = \{\dim [Q_x(k,2)]\}^{N/x} = \left[\sum_{j_2 = 0}^{k/2}
{ x \choose {j_2}} {{x-j_2} \choose {k-2j_2}} \right]^{N/x},
\end{eqnarray*}
where $j_2$ indexes the number of sites with two electrons per
site and is allowed to range from $0$ to $N/2$ for $x$ even, or
$0$ to $(N-1)/2$ for $x$ odd.  It is straightforward to show that
the dimensionality of each partition is maximized when $k=x$, i.e.
the number of particles per element is equal to the number of
sites per element.  The dimensionality in this case is shown by
the crosses, $\times$, in Fig. \ref{fig:CompareDim}, which is
significantly larger than the case where at most one particle is
allowed per site.

Another alternative is where we have an extra quantum number (for
example spin), which would double the number of accessible quantum
states.  In this case the system returns to the standard Fermionic
case and the dimensionality becomes simply
\begin{eqnarray*}
{\cal D} = {{2x} \choose k}^{N/x}.
\end{eqnarray*}
This dimensionality is shown by the asterisks, $\ast$, in Fig.
\ref{fig:CompareDim} which represents the largest dimensionality
of all the cases under consideration.

In general, we note that if we allow the number of particles per
site to be at most $z$, then the total Hilbert-space
dimensionality is
\begin{widetext}
\begin{eqnarray*}
\{\dim [Q_x(k,z)]\}^{N/x} = \left[\sum_{j_z = 0}^{x/z} {x \choose
{j_z}}
  \sum_{j_{z-1}=0}^{(k-zj_z)/(z-1)} {{x-j_z} \choose {j_{z-1}}}
  \cdots
  \sum_{j_2 = 0}^{(k - \sum_{r=3}^z r j_r)/2}
  {{x- \sum_{r=3}^z j_r} \choose {j_2}}
  {{x- \sum_{r=2}^z j_r} \choose
  {k- \sum_{r=2}^z r j_r}}\right]^{N/x}.
\end{eqnarray*}
\end{widetext}

Note that if there is no restriction on the number of particles
per site we retrieve the Bosonic case.  We would then be looking
to extend familiar bosonic quantum computing schemes to the
multiple-particle limit, for example linear optical
\cite{bib:KLM}, nonlinear optical \cite{bib:QNLO} or
superconducting \cite{bib:CooperPairBox} quantum computing
schemes.  The dimensionality is considerably simplified and is
\cite{bib:Blume-Kohout2002,bib:MayerText}
\begin{eqnarray*}
\dim [Q_x(k,k)] = {{x+k-1} \choose {x-1}}.
\end{eqnarray*}
This dimensionality can greatly exceed the fermionic case, for the
same number of accessible quantum states, for a large enough
number of particles.  Furthermore this dimensionality
monotonically increases with increasing particle number.  This
suggests that a quantum computing architecture based around bosons
may be easier to upgrade than any other architecture.

We note in closing that there is far more to quantum computers
than optimizing the Hilbert space dimension.  Here we have not
entered into issues of operational complexity, which will be
specific to particular architectures, as well as decoherence which
is expected to differ substantially between implementations.
However, comparing qudit implementations, a system comprising
qubits alone will maximize the amount of information that can be
obtained in a single measurement step. This is because an optimum
measurement would be to measure the state of every quantum
particle.  For qubits, that would yield $N/2$ values; for qutrits,
it would only yield $N/3$ values.  As a rough measure, we may say
that to equalize the amount of information gained about the state
of the computer, we would need to perform $(N/2)/(N/3) = 1.5$
times as many measurements on a qutrit-based quantum computer as a
qubit-based quantum computer. The dimensionality of the qutrit
Hilbert space exceeds that of the qubit Hilbert space by $1.5$
when
\begin{eqnarray*}
\exp\left[N\left(\frac{\ln 3}{3} - \frac{\ln 2}{2} \right)\right]
> 1.5,
\Rightarrow N > 20.65. \\
\end{eqnarray*}
Therefore when the number of quantum sites exceeds 20.65, the
increase in dimensionality of the Hilbert space should more than
compensate for the increased measurement complexity.

We have shown, using basic arguments, that for a quantum computer
with a finite number of distinguishable states to be shared
between a finite number of quantum particles arranged in qudits,
the Hilbert space is maximized when the system is partitioned into
the largest possible number of qutrits.  Given that quantum
computers suffer many limitations, this kind of optimization for a
given geometry may make the difference between a practical and
impractical implementation of quantum computing, despite the fact
that both qubit and qutrit based quantum computers are scalable.
The Hilbert-space dimensionality can be increased substantially by
both relaxing the requirement that the particles be partitioned
into qudits, and by implementing what we have termed a packing
geometry where the particles are allowed to be present in any
site, up to some maximum which would be determined by the physical
properties of the quantum system.

ADG would like to acknowledge helpful discussions with Dr Stephen
Bartlett of Maquarie University, Australia, and Dr Andrew White of
the University of Queensland, Australia. SGS would like to
acknowledge financial support from the CMI project on Quantum
Information.  This work was supported by the Australian Research
Council.


\end{document}